\documentclass{article}
\usepackage{amsmath}
\usepackage{amssymb}
\usepackage[numbers,square]{natbib}
\usepackage{hyperref}
\hypersetup{
	colorlinks=true,
	linkcolor=blue,
	filecolor=magenta,      
	urlcolor=cyan
}
\usepackage{graphicx}
\usepackage[margin=1.25in]{geometry}
\usepackage{booktabs}
\usepackage{authblk}

\begin{document}
	
	\title{A scalable, synergy-first backbone decomposition of higher-order structures in complex systems.} 
	\author[1,2,*]{Thomas F. Varley}
	
	\affil[1]{Vermont Complex Systems Center, University of Vermont, Burlington, VT, 05405}
	\affil[2]{Department of Computer Science, University of Vermont, Burlington, VT, 05405}
	\affil[*]{tfvarley@uvm.edu}
	
	\maketitle
	
	\begin{abstract}
		Since its introduction in 2011 by Williams and Beer, the partial information decomposition (PID) has triggered an explosion of interest in the field of multivariate information theory and the study of emergent, higher-order (``synergistic") interactions in complex systems. Despite its considerable power however, the PID has a number of limitations that restrict its general applicability: it scales very poorly with system size, and the zoo of different information ``atoms" rapidly becomes uninterpretable for even small systems. Finally, the standard approach to decomposition hinges on a definition of ``redundancy", and leaves synergy only vaguely defined as ``that information not redundant." Since then, other heuristic measures, such as the O-information have been introduced, although these measures typically only provided a summary statistic of redundancy/synergy dominance, rather than direct insight into the synergy itself. To address this issue, we present an alternative decomposition that is synergy-first, scales much more gracefully than the antichain lattice, and has a straightforward interpretation. Our approach defines synergy as that information encoded in the joint state of a set of elements that would be lost following the minimally invasive perturbation on any single element. By generalizing this idea to sets of elements, we construct a totally ordered ``backbone" partial synergy atoms, each one representing successively less fragile information that sweeps systems scales. We begin by demonstrating this decomposition on the joint entropy before showing that it can be generalized to the Kullback-Leibler divergence, and by extension, to the total correlation and the single-target mutual information (thus recovering a ``backbone" PID). Finally, we show that this approach can be used to decompose higher-order interactions beyond just information theory: we demonstrate this by showing how synergistic combinations of pairwise edges in a complex network supports signal communicability and global integration. We conclude by discussing how this perspective on synergistic structure (information-based or otherwise) can deepen our understanding of part-whole relationships in complex systems.
	\end{abstract}
	
	\section*{Introduction}
	
	To what degree can a ``whole" complex system be said to be ``greater than the sum of its parts?" This is one of the core questions in modern complexity theory, as the emergence of higher-order phenomena from the interactions between lower-order elements is a defining feature of complex systems throughout the natural and artificial worlds \cite{artime_origin_2022}. Some of the most promising tools for interrogating these emergent phenomena come from information theory, which provides a mathematically rigorous formal language with which to explore part-whole relationships in multivariate systems \cite{varley_information_2023}. In the context of information theory, information that is present in the joint state of the whole, but none of the parts is known as ``synergy", and has been proposed as a formal statistic of emergent behaviours in complex systems \cite{mediano_greater_2022,varley_emergence_2022,varley_flickering_2023}. Synergy has also been found to be ubiquitous in natural and artificial systems, having been observed in networks of cortical neurons \cite{newman_revealing_2022}, whole-brain fMRI dynamics \cite{varley_multivariate_2023,luppi_synergistic_2022}, climatological systems \cite{goodwell_temporal_2017}, interactions between social identities and life-outcomes (often called ``intersectionality") \cite{varley_untangling_2022}, and heart-rate dynamics \cite{pinto_multiscale_2022}. Furthermore, synergy has been proposed to inform about clinically-meaningful processes such as ageing \cite{gatica_high-order_2021}, traumatic brain injury \cite{luppi_reduced_2023}, and the actions of anaesthetic drugs \cite{luppi_synergistic_2023}. This list is non-exhaustive. 
	
	Despite the clear significance of synergistic information in a variety of fields and contexts, the mathematical tools for exploring the space of synergistic dependencies remain limited. There are currently two broad families of methods: those based on the partial information decomposition (PID) \cite{williams_nonnegative_2010,gutknecht_bits_2021}, and those based on the O-information \cite{rosas_quantifying_2019} and other multivariate extensions of mutual information \cite{james_anatomy_2011}. While both families of methods are powerful, and have proven to be useful in many contexts, they suffer from limitations that restrict their applicability in some cases. The PID-like approaches (such as the partial entropy decomposition \cite{ince_partial_2017,finn_generalised_2020}, the integrated information decomposition \cite{rosas_reconciling_2020}, and the generalized information decomposition \cite{varley_generalized_2024}) provide a ``complete" picture of the whole structure of multivariate information, in the form of a partially ordered lattice of ``atomic" interactions between elements. Unfortunately, the size of the lattice (and consequently, the number of operations required to compute the synergy) balloons super-exponentially: the number of information atoms in a system with $k$ elements is the $k^{\textnormal{th}}$ Dedekind number minus two. It is simply not feasible to apply any PID-like analysis to a system with more than five elements. The same system grows, the atoms themselves becomes largely uninterpretable: it's hard to imagine the study where we care deeply about that information that could be learned by observing ($X_1$ and $X_2$) or ($X_1$ and $X_3$ and $X_7$) or $X_4$ or ($X_5$ and $X_6$ and $X_7$), let alone how it differs from the information that could be learned by observing ($X_3$ and $X_2$) or ($X_1$ and $X_3$ and $X_7$) or $X_4$ or ($X_1$ and $X_6$ and $X_7$). Finally, and perhaps most problematically for the study of synergy, almost all PIDs take a redundancy-first approach, and synergy is only implicitly defined as ``that information left over when all the redundancies have been partialed out" (although a notable exception is the synergistic disclosure framework, discussed below \cite{rosas_operational_2020}). 
	
	The O-information family of measures are more heuristic. Rather than provide a complete picture of the structure of multivariate information, they provide insight into whether a system is \textit{dominated} by redundancy or synergy. The O-information and related measures scale gracefully with system size, but do not provide much of a map of how synergies of different orders are present in the system. Furthermore, the O-information generally takes a very conservative definition of synergy and a very liberal definition of redundancy: for a $k$-element system, the synergy is only that information that is in all $k$ elements, and any information in any set of elements less than $k$ gets classified as ``redundancy", even if it is information only accessible in the joint state of many elements \cite{varley_generalized_2024}. A more liberal definition of synergy that allows researchers to consider different orders of synergy (beyond just the topmost) would be useful in many contexts. 
	
	In this work, I introduce a novel, heuristic approach to analysing synergistic information in complex systems. This approach is synergy-first (beginning with an intuitive definition of what it means for information to be synergistic), scales much more manageably than the PID, but also provides insights into the presence (or absence) of synergy at every scale. This approach is localizable \cite{lizier_local_2013}, and can be applied to a variety of information-theoretic quantities, including (but not limited to) the entropy (in which case the decomposition is strictly non-negative), the single-target mutual information, the Kullback-Leibler divergance, and the total correlation. Finally, I show that the logic of this decomposition can be applied outside the specific world of information theory to interrogate the structure of complex systems using other methods entirely. 
	
	\section*{What is synergistic information?}
	
	Above, we defined ``synergy" intuitively as ``that information that is in the whole but none of the parts." How can we make this definition formal? Consider an information source \textbf{X} comprised of $k$ channels (for conceptual simplicity, we will assume everything is discrete): $\textbf{X} = \{X_1,\ldots,X_k\}$. At every moment, you observe some specific realization of $\textbf{X} = \textbf{x}$, which is drawn from a joint probability distribution $\mathbb{P}(\textbf{X})$ on the support set $\boldsymbol{\mathcal{X}}$. The total Shannon information content (or local entropy) of \textbf{x} is given by the local entropy: $h(\textbf{x})=-\log \mathbb{P}(\textbf{x})$. While there has been considerable prior work on redundancy-based approaches for decomposing $h(\textbf{x})$ (for three options see: \cite{ince_partial_2017}, \cite{finn_generalised_2020}, and \cite{varley_partial_2023}), here we will take a synergy-first approach and say:
	
	\begin{quotation}
		\textit{The synergistic information in $\textbf{x}=\{x_1,\ldots,x_k\}$ is that information which is only disclosed if $X_1=x_1\land\ldots\land X_k=x_k$ are observed simultaneously, and cannot be disclosed by any subset of \textbf{x} of size $k-1$ or less.}
	\end{quotation}

	What are the implications of this definition? Suppose that our goal was to infer the state of \textbf{X}. If all the channels were functioning correctly, this would be trivial: we would just have to look and see that \textbf{X}=\textbf{x}. However, if one of the channels were to fail, the problem of inferring \textbf{x} becomes necessarily harder. Assuming we have access to the joint distribution $\mathbb{P}(\textbf{X})$, it may be possible to use correlations between the various $X_i$ (or higher-order correlations between various subsets of \textbf{X}) to recover \textbf{x}, fully or partially. In the context of the PID, this might be called the ``redundancy". We are interested in assessing the synergy directly though, and in our context the synergy is that information about \textbf{x} that \textit{cannot} be recovered without knowing the state of every channel simultaneously, and which is destroyed when access to any one channel is lost.  
	
	Consider, if the channel between the source and the observer were to break and stop returning a value for $X_1$, then presumably the synergistic information that could have been learned from $\{x_1,\ldots,x_k\}$ will be destroyed, since only information that could be learned from $\{x_2,\ldots,x_k\}$ is still accessible. In fact, this synergistic information would be destroyed by \textit{any} single failure of \textit{any} $X_i$, as that would be enough to disrupt the requirement that $X_1=x_1\land\ldots\land X_k=x_k$.
	
	Inspired by Makkeh et al.,'s work linking redundant and synergistic information to logical connectives \cite{makkeh_introducing_2021}, we could rephrase the definition of synergy, then as:
	
	\begin{quotation}
		\textit{The synergistic information in $\textbf{x}=\{x_1,\ldots,x_k\}$ is that information that would be destroyed if $X_1$ alone failed \textbf{or} $X_2$ alone failed ... \textbf{or} $X_k$ alone failed.} 
	\end{quotation}

	It doesn't matter which $X_i$ fails: the failure of any one alone is enough to destroy the synergy. This leads us to the straightforward formal definition:
	
	\begin{align}
	h^{syn}(\textbf{x}) &= \min_{i}\bigg[h(\textbf{x}) - h(\textbf{x}^{-i})\bigg] \\
	&= \min_{i} \bigg[ h(x_i|\textbf{x}^{-i})\bigg ]
	\end{align}
	
	Where $\textbf{x}^{-i}$ refers to the joint state of every $x_j$ in \textbf{x} excluding $x_i$. This is the amount of information that you would lose if any element failed, regardless of which one it is. The value of $h^{syn}(\textbf{x})$ is that portion of the information in \textbf{x} that depends on the joint state of all elements and is fragile to the failure of any one. It's implicitly defined as a local value, but the expected synergy can be computed in the usual way:
	
	\begin{equation}
	H^{syn}(\textbf{X}) = \mathbb{E}_{\mathbb{P}(\textbf{X})}\bigg[h^{syn}(\textbf{x}) \bigg]
	\end{equation} 
	
	\subsection*{The $\boldsymbol{\alpha}$-synergy backbone}
	
	\begin{figure}
		\centering
		\includegraphics[width=0.9\textwidth]{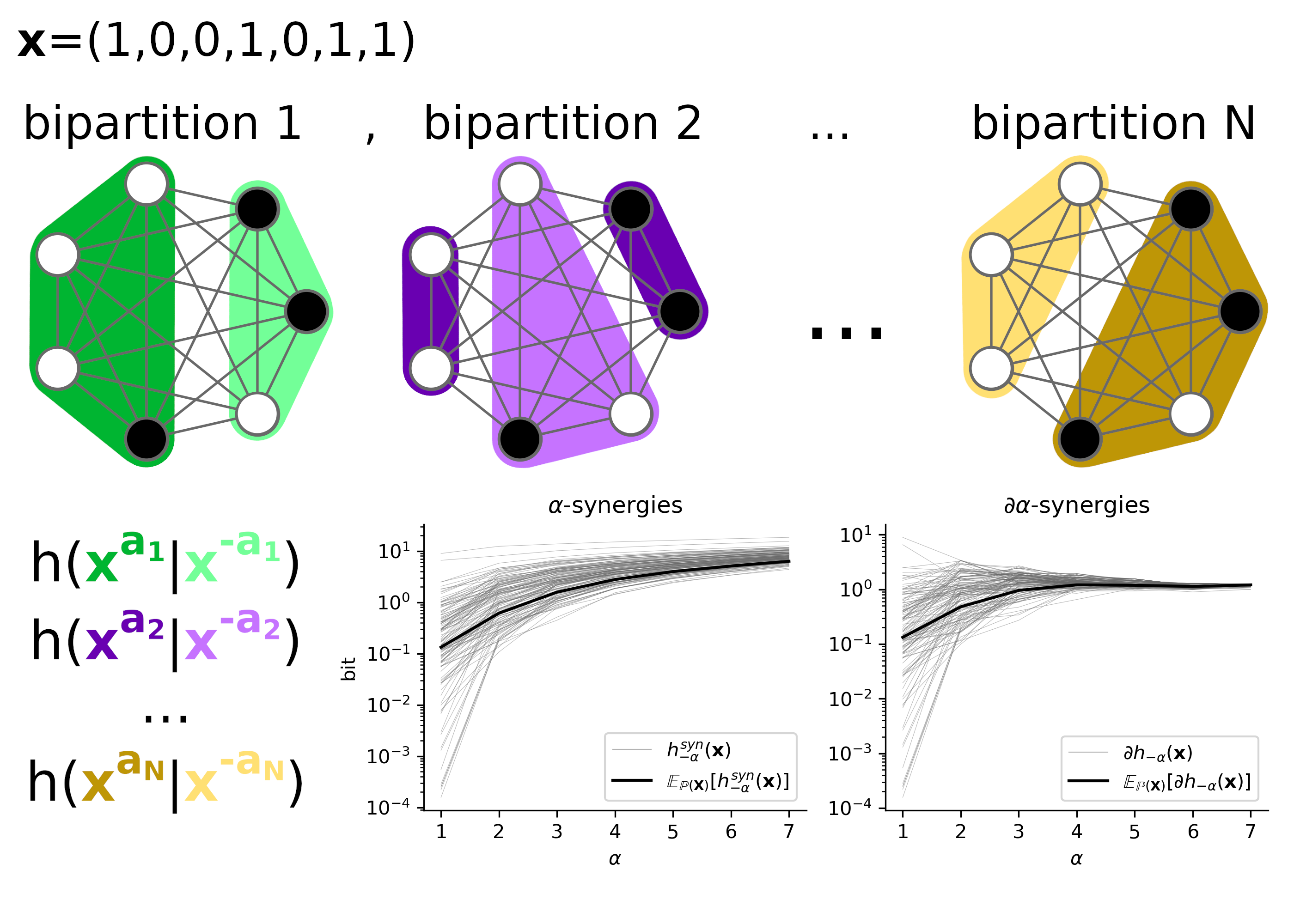}
		\caption{\textbf{The bipartition approach to local $\boldsymbol{\alpha}$-synergistic entropy.Top row:} For a seven-element system \textbf{X} in a particular state (\textbf{x}=(1,0,0,1,0,1,1)), the $\alpha$-synergistic entropy decomposition requires partitioning \textbf{x} into all possible subsets of size $\alpha$, and finding the minimum value of $h(\textbf{x}^{\textbf{a}}|\textbf{x}^{-\textbf{a}})$. \textbf{Bottom row:} by sweeping all scales $1\ldots k$, it is possible to construct a hierarchical picture of how fragile synergy is distributed over scales.}
		\label{fig:explainer}
	\end{figure}

	The information that would be destroyed by any single failure is clearly the most fragile information. All the remaining information must be robust to at least one failure. We can define the dual measure of robustness as:
	
	\begin{align}
	h^{rbst}(\textbf{x}) = h(\textbf{x}) - h^{syn}(\textbf{x})
	\end{align}
	
	This notion of robustness has some parallels to the notion of redundancy in the partial information decomposition, although we will not dwell on it here. One might naturally ask: what if more than one element fails? What information is robust to the failure of any two elements? What information is synergistic, not in the joint-state of all $k$ elements, but any subset of $k-1$ elements? This is the basic logic of the synergy ``backbone" decomposition. The first thing we will do is re-name the measure $h^{syn}(\textbf{x})$ to $h_{1}^{syn}(\textbf{x})$ to indicate that it is the information that is destroyed by a single failure. But you could easily have the $2$-synergy:
	
	\begin{align}
	h_{2}^{syn}(\textbf{x}) &= \min_{ij}\bigg[h(\textbf{x}) - h(\textbf{x}^{-\{ij\}})\bigg] \\
	&= \min_{ij}\bigg[h(\textbf{x}^{ij}|\textbf{x}^{-\{ij\}})\bigg]
	\end{align}
 
	This is the information in \textbf{x} that would be lost if two channels, $X_i$ and $X_j$, failed, regardless of which two failed specifically. Returning to the link with logical connectives, it is the information that would be lost if $X_1$ \textit{and} $X_2$ failed \textbf{or} $X_1$ \textit{and} $X_3$ failed \textbf{or} $\ldots$ and so on. In general, I can define the $\alpha$-synergy as the information destroyed when any set of $\alpha$ elements fails:
	
	\begin{align}
	h_{\alpha}^{syn}(\textbf{x}) &= \min_{\substack{\textbf{a}\subset \{k\} \\ |\textbf{a}|=\alpha}}\bigg[h(\textbf{x}) - h(\textbf{x}^{-\textbf{a}})\bigg] \\ 
	&= \min_{\substack{\textbf{a}\subset \{k\} \\  |\textbf{a}|=\alpha }}\bigg[h(\textbf{x}^{\textbf{a}}|\textbf{x}^{-\textbf{a}})\bigg] \label{eq:symmetry}
	\end{align} 
	
	Where $\{k\}$ is shorthand for the set $\{0,\ldots,k\}$. It is clear that the $\alpha$-synergistic entropy function is non-negative (due to the non-negativity of local conditional entropy), and it is monotonically increasing on the interval $\{0,\ldots,k\}$ (for proof, see Appendix). 
	
	The combination of non-negativity and monotonicity imply that the $\alpha$-synergy function can decompose $h(\textbf{x})$ into a set of non-negative partial synergies that account for the information intrinsic to each scale. Recall that the 1-synergy is the information in \textbf{x} that would be lost if any one element failed, regardless of which specific element it actually is. Similarly, the 2-synergy is the information in \textbf{x} that would be lost if any two elements failed. However, the information in the 2-synergy must, definitionally, contain the 1-synergy, since the two elements failing simultaneously must include at least as much fragile information as the 1-synergy. In the same vein as the classic partial information decomposition, we can propose a bootstrapping approach to partial out the synergistic information intrinsic to each scale by defining the $\alpha$-partial synergy function:
	
	\begin{equation}
	\partial h_{-\alpha}(\textbf{x}) = h_{\alpha}^{syn}(\textbf{x}) - \sum_{\beta < \alpha}\partial h_{\beta}(\textbf{x})
	\end{equation} 
	
	We relax the use of the \textit{syn} superscript for the partial entropy function, as it is assumed that any partial term is being computed with the $\alpha$-synergy function. This completes the decomposition of the local entropy:
	
	\begin{equation}
	h(\textbf{x}) = \sum_{i=1}^k \partial h_{i}^{syn}(\textbf{x})
	\end{equation}
	
	As with the $\alpha$-synergy function, an expected value over all realizations can be computed in the usual way:
	
	\begin{equation}
	\partial H_{-\alpha}(\textbf{X})= \mathbb{E}_\textbf{X}\bigg[\partial h_{\alpha}(\textbf{x})\bigg]
	\end{equation}
	
	This spectrum of $\alpha$-partial synergies forms the ``backbone" decomposition of the entropy of \textbf{X}. In contrast to the redundancy-based partial entropy decomposition, which will produce a large lattice of different ``atomic" entropies, this backbone decomposition only produces $k$ values, arranged in order of increasingly robust synergies from 1 to $k$. 
	
	\subsubsection*{Discrete vs. differential entropy}
	
	In this paper, we assume that all random variables are discrete, with finite-sized support sets. This is typical in information theory, which is most naturally at home among discrete random variables. However, the natural world is not discrete (at least, not at the levels that most scientists are interested in), and so the question of how to apply a given information-theoretic analysis to continuous random variables inevitably comes up. The generalization of the discrete Shannon entropy to the continuous, differential entropy is reasonably straightforward, and closed-form estimators exist for a variety of parametric distributions \cite{cover_elements_2012}. The most commonly seen differential entropy estimators are for univariate and multivariate Gaussian distributions, where the local entropy $h^{\mathcal{N}}(\textbf{x})$ can be computed as:
	
	\begin{equation}
	h^{\mathcal{N}}(\textbf{x}) = -\ln\bigg[ \frac{\exp\left(-\frac 1 2 \left({\mathbf x} - {\boldsymbol\mu}\right)^\mathrm{T}{\boldsymbol\Sigma}^{-1}\left({\mathbf x}-{\boldsymbol\mu}\right)\right)}{\sqrt{(2\pi)^k |\boldsymbol\Sigma|}}\bigg]
	\end{equation}
	
	Where $\boldsymbol{\mu}$ is the vector of marginal means, $\boldsymbol{\Sigma}$ is the covariance matrix of \textbf{X}, $k$ is the dimensionality of \textbf{x}, $\boldsymbol{\Sigma}^{-1}$ is the matrix inverse of $\boldsymbol{\Sigma}$, and $|\boldsymbol{\Sigma}|$ the determinant. The local differential entropy can be used to build up all the usual information-theoretic measures, both local and expected (for a review, see \cite{lizier_jidt_2014}), although it is important to note that the differential entropy differs from the discrete entropy in several key ways, that could complicate the analyses presented here. The most significant is that the local differential entropy can be negative, since the Gaussian probability density can be greater than unity (depending on the specific values of $\boldsymbol{\mu}$ and $\boldsymbol{\Sigma})$. Since much of the interpretation of the entropy relies on the non-negativity of joint and conditional entropies, this presents a conceptual, if not a practical, problem for attempts to generalize discrete information-theoretic tools to continuous random variables. Despite this, Gaussian estimators of differential entropy are common in complex systems science, particular in neuroimaging, where continuous-valued time series are common (\cite{varley_multivariate_2023,luppi_synergistic_2022,tononi_measure_1994}), and on some cases, exact, closed-form Gaussian estimators of previously-discrete measures can be derived analytically \cite{barrett_exploration_2015,kay_exact_2018}. Future work understanding the behaviour of the $\alpha$-synergy decomposition for differential entropies will help deepen our understanding of higher-order information sharing in many types of complex systems. 
	
	\subsubsection*{Alternative formulation}
	\label{sec:alternate_forms}
	Here, we use the minimum information loss over all possible sets of $\alpha$ failures to specify the synergy at a given scale, to match the intuition that the $\alpha$-synergy is a kind of intersection: that information lost common to all subsets of size $\alpha$. However, other formulations may also be useful, for example, instead of computing $\min_{\textbf{a}}\big[h(\textbf{x}) - h(\textbf{x}^{-\textbf{a}})\big]$, one might compute the average value of $h(\textbf{x})-h(\textbf{x}^{-\textbf{a}})$ over all $\textbf{x}^{-\textbf{a}}$. In this case, the interpretation is slightly different: rather than asking ``what information is guaranteed to be lost if any set of $\alpha$ elements failed", it asks ``how much information would the observer expect to lose if any set of $\alpha$ elements failed." These are subtlety different definitions of synergy (perhaps analogous to the problem fo multiple redundancy functions in the PID), although the behaviour of the resulting $\alpha$-synergy functions is largely the same (the average case is also non-negative and monotonically increasing, see Appendix for proof). This alternative formulation more closely resembles the Tononi-Sporns-Edelman (TSE) complexity \cite{tononi_measure_1994}, discussed in detail below.
	
	Alternately, one could adopt a worst-case-scenario perspective and replace the min function with a max function. Then the synergy becomes a measure of the most fragile information contained in any combination of elements. Like the average approach, the use of the max function also preserves the required behaviour of the $\alpha$-synergy function, albeit with a different interpretations (see Appendix). Different specific cases and contexts may call for different formulations. 
	
	\subsection*{Computing the $\boldsymbol{\alpha}$-synergy decomposition for large systems}
	
	While the $\alpha$-synergy decomposition scales more managably than the PID, it can still get unwieldy for large systems. While the runtime complexity of the PID grows with the Dedekind numbers, the complexity of the $\alpha$-synergy decomposition grows with the Bell numbers. As we can see in Eq. \ref{eq:symmetry}, the $\alpha$-synergy implicitly requires computing the conditional entropy every subset of \textbf{x} of size $\alpha$ and its complement, over the range of integers $1\ldots k$, this works out to every possible subset of \textbf{x}. For a system with eight elements, that is $\approx4140$ bipartitions to test (although compare that to the number of partial information atoms in an eight element system: 56,130,437,228,687,557,907,788). 
	
	The problem of finding the bipartition of size $\alpha$ that minimizes the local conditional entropy is closely related to the problem of finding the \textit{minimum information bipartition} in the context of integrated information theory \cite{tononi_measuring_2003}, and has been explored extensively over the last two decades. In cases where a system is too large for the minimum local conditional entropy to be computed directly, a number of heuristic options are available. One is simply random sampling, recording the minimally entropic bipartition as it goes. Alternately, one might consider an optimization approach, such as simulated annealing, or a more bespoke approach like Queyranne’s algorithm \cite{kitazono_efficient_2018}, to find the winning bipartition. If one uses the alternative formulation based on expected values over bipartitions discussed above, the approach is somewhat easier, as it requires merely doing a large number of sample of possible $\textbf{x}^{\textbf{a}}$, as was done in \cite{varley_multivariate_2023,varley_evolving_2024} with the TSE complexity, rather than optimizing a particular minimum value. 
	
	If taking an optimization or random-sampling approach to find the $\alpha$-synergy for a large system, care must be taken that the monotonicity criteria is not violated. If the algorithm fails to find the global minima for each value of $\alpha$, it is possible (although unlikely) that the ``winning" bipartition for scale $\beta$ is less than the ``winning" bipartition for scale $\alpha$, even if $\alpha < \beta$. This will result in negative partial synergistic entropy atoms, compromising the interpretation of the decomposition. Maintaining the non-negativity of partial entropy atoms becomes important when generalizing the synergistic entropy decomposition to other measures, such as the Kullback-Leibler divergence. 
	
	\section*{Extending the $\boldsymbol{\alpha}$-synergy decomposition.}
	
	The Shannon entropy forms the foundation of a large number of more complex information-theoretic measures and by decomposing the component entropies of these measures, we can generalize the $\alpha$-synergistic entropy decomposition to formalize notions of synergy in a variety of other  measures. This section with largely recapitulate the logic in \cite{varley_generalized_2024}: first showing how the local entropy decomposition induces a more general decomposition of the Kullback-Leibler divergence, from which it is then possible to construct $\alpha$-synergy decompositions of the negentropy, total correlation, and single-target mutual information. By the end, we will have seen that the $\alpha$-synergy decomposition can be deployed in almost every situation that the ``full" partial information decomposition is deployed in. 
	
	\subsection*{The Kullback-Leibler divergence}
	Recently Varley showed that a partial entropy decomposition can be used to construct a decomposition of the Kullback-Leibler divergence \cite{varley_generalized_2024}. The divergence  quanitifies the information gained when one updates from a prior set of beliefs (typically indicated by $\mathbb{Q}(\textbf{X})$) to a new set of posterior beliefs (typically indexed by $\mathbb{P}(\textbf{X})$). The typical formal presentation is:
	
	\begin{align}
	D^{\mathbb{P}||\mathbb{Q}}(\textbf{X}) &= \sum_{\textbf{x}\in\mathcal{\textbf{X}}}\mathbb{P}(\textbf{x})\log\frac{\mathbb{P}(\textbf{x})}{\mathbb{Q}(\textbf{x})} \\
	&= \mathbb{E}_{\mathbb{P}(\textbf{X})}\bigg[\log\frac{\mathbb{P}(\textbf{x})}{\mathbb{Q}(\textbf{x})}\bigg]
	\end{align}
	
	Which is read as ``the divergence of $\mathbb{P}$ from $\mathbb{Q}$." This can be re-written in purely information theoretic terms to show: 
	
	\begin{equation}
	D^{\mathbb{P}||\mathbb{Q}}(\textbf{X}) = \mathbb{E}_{\mathbb{P}(\textbf{X})}\bigg[{h^{\mathbb{Q}}(\textbf{x}}) - h^{\mathbb{P}}(\textbf{x}) \bigg]
	\end{equation}
	
	Where $h^{\mathbb{Q}}(\textbf{x})$ indicates that the local entropy of \textbf{x} is computed from the distribution $\mathbb{Q}(\textbf{x})$. The $\alpha$-partial synergy decomposition allows us to decompose $h^{\mathbb{Q}}(\textbf{x})$ and $h^{\mathbb{P}}(\textbf{x})$ into $k$ pairs of non-negative partial synergy atoms. The difference between the elements of each pair is the local $\alpha$-synergistic divergence: 
	
	\begin{equation}
	\partial d^{\mathbb{P}||\mathbb{Q}}_{\alpha}(\textbf{x}) = \partial h^{\mathbb{Q}}_{\alpha}(\textbf{x}) - \partial h^{\mathbb{P}}_{\alpha}(\textbf{x}) 
	\end{equation}
	
	Unlike the partial entropy atoms, the partial divergence atoms are not non-negative, however the negative values are easily interpretable: $\partial d^{\mathbb{P}||\mathbb{Q}}_{\alpha}(\textbf{x}) < 0$ simply implies that there is \textit{more} $\alpha$-synergistic surprise in \textbf{x} when we believe our posterior $\mathbb{P}$ than when we believed our prior $\mathbb{Q}$. The local $\alpha$-synergistic divergence can be aggregated into an expected value:
	
	\begin{equation}
	\partial D^{\mathbb{P}||\mathbb{Q}}_{-\alpha}(\textbf{X}) = \mathbb{E}_{\mathbb{P}(\textbf{X})}\bigg[\partial d^{\mathbb{P}||\mathbb{Q}}_{-\alpha}(\textbf{x})\bigg]
	\end{equation}
	
	Which is also not non-negative, although again, this is not particularly mysterious: it just means that, on average, here was more $\alpha$-synergistic surprise in the prior than the posterior at the scale given by $\alpha$. This completes the decomposition of the expected Kullback-Leibler divergence:
	
	\begin{equation}
	D^{\mathbb{P}||\mathbb{Q}}(\textbf{X}) = \sum_{i=1}^{k}\partial D^{\mathbb{P}||\mathbb{Q}}_{-i}(\textbf{X})
	\end{equation}
	
	A variety of different measures can be written in terms of the Kullback-Leibler divergence, and are now fair game for $\alpha$-synergy decomposition. Below we will discuss three: the negentropy and the total correlation (two measures of divergence from maximum-entropy independence) and the single-target mutual information (the original impetus for the partial information decomposition). 
	
	\subsection*{Special cases of the Kullback-Leibler divergence}
	
	The Kullback-Leibler divergence defines a large number of different information-theoretic metrics, which can be decomposed into various levels of $\alpha$-synergy. An exhaustive discussion of all of them is beyond the scope of this manuscript, but two areas that warrant further exploration are the negentropy \cite{brillouin_negentropy_2004} and the total correlation \cite{watanabe_information_1960}. 
	
	Both are measures of ``deviation from independence", although they define ``independence" in different ways. For a random variable \textbf{X} with probability distribution $\mathbb{P}(\textbf{X})$, the negentropy is:
	
	\begin{equation}
	N(\textbf{X}) = \sum_{\textbf{x}\in\mathcal{\textbf{X}}}\mathbb{P}(\textbf{x})\log\frac{\mathbb{P}(\textbf{x})}{\mathbb{U}(\textbf{x})} 
	\end{equation} 
	
	Where $\mathbb{U}(\textbf{X})$ is the uniform probability distribution: for all \textbf{x} in $\mathcal{X}$, $\mathbb{U}(\textbf{x})=1/|\mathcal{X}|$. The uniform distribution is maximally entropic, it has no ``structure", and under the uniform distribution every subset of \textbf{X} is also maximum-entropy. The negentropy, then, quantifies something like ``how different is \textbf{X} from a kind of ideal gas?" Alternately, Rosas et al., describe it as ``the information about the system that is contained in its statistics." \cite{rosas_understanding_2016}. 
	
	A closely-related measure, the total correlation, is also a measure of independence, but it preserves the first-order marginal structure:
	
	\begin{equation}
	TC(\textbf{X}) = \sum_{\textbf{x}\in\boldsymbol{\mathcal{X}}}\mathbb{P}(\textbf{x})\log\frac{\mathbb{P}(\textbf{x})}{\prod_{i=1}^{k}\mathbb{P}(x_i)}
	\end{equation}
	
	The total correlation is zero if every $x_i\in\textbf{X}$ is independent of every other $x_j\in\textbf{X}$, even if the individual $x_i$ themselves are not maximum entropy. In some sense it is a more conservative definition of ``deviation from independence."
	
	Both the negentropy and the total correlation quantify how much information is gained when we update our beliefs from a prior that \textbf{X}'s statistics are ``unstructured" to a posterior of the true statistics. They define what it means to be ``structured" in different ways (maximum entropy versus marginally independent), but the basic intuition is the same. When we apply the $\alpha$-synergy decomposition, what we get is a partition of the information gain into $k$ atoms, each of which depends on progressively lower-order combinations of elements to remain stable to perturbation.
	
	\subsubsection*{Examples}
	
	To demonstrate this in more detail, consider the logical XOR function: a typical toy model for demonstrating ``synergy" in discrete systems. We will say that $X_1\oplus X_2 = Y$, with a maximum entropy distribution on $X_1,X_2$. In the case of the XOR, the total correlation and the negentropy are the same, so we will refer to the synergistic negentropy atoms for preference. 
	
	\begin{table}[h!]
		\centering
		\begin{tabular}{@{}ccccc@{}}
			\toprule
			$\mathbb{P}(x_1,x_2,y)$ & $(x_1,x_2,y)$ & $\partial n_{1}$ & $\partial n_{2}$ & $\partial n_{3}$ \\ 
			\midrule
			1/4 & (0,0,0) & 1 bit & 0 bit & 0 bit \\
			1/4 & (0,1,1) & 1 bit & 0 bit & 0 bit \\
			1/4 & (1,0,1) & 1 bit & 0 bit & 0 bit \\
			1/4 & (1,1,0) & 1 bit & 0 bit & 0 bit \\ \bottomrule
		\end{tabular}
		\caption{\textbf{The} $\boldsymbol{\alpha} $\textbf{-synergy decomposition of the negentropy of the logical-XOR gate.}}
		\label{tab:neg_xor}
	\end{table}
	
	For each of the possible states, all of the partial negentropy is in the $1-synergy$: recall from the argument above that $\partial n_{\alpha}(\textbf{x})$ is equal to $\partial h^{\mathbb{U}}_{\alpha}(\textbf{x}) - \partial h^{\mathbb{P}}_{\alpha}(\textbf{x})$. The positive value of 1 bit shows that, under our prior of maximum entropy, there is 1 bit of synergistic uncertainty in the joint state of $x_1,x_2,y$, but after updating our beliefs to the true distribution, that bit is resolved. After losing any single element, all of the remaining subsets are themselves maximum entropy, and so are not different from the prior in any way.
	
	We can also consider the decomposition of the logical-AND system, with $X_1 \land X_2 = Y$. Unlike the logical-XOR system, the logical-AND system is not symmetric, and will show different behaviours for different states. 
	
	\begin{table}[h!]
	\centering
		\begin{tabular}{@{}ccccc@{}}
			\toprule
			$\mathbb{P}(x_1,x_2,y)$ & $(x_1,x_2,y)$ & $\partial n_{1}$ & $\partial n_{2}$ & $\partial n_{3}$ \\ 
			\midrule
			1/4 & (0,0,0) & 1 bit & 0 bit & 0 bit \\
			1/4 & (0,1,0) & 1 bit & 0 bit & 0 bit \\
			1/4 & (1,0,0) & 1 bit & 0 bit & 0 bit \\
			1/4 & (1,1,1) & 1 bit & 1 bit & -1 bit \\ \bottomrule
		\end{tabular}
		\caption{\textbf{The} $\boldsymbol{\alpha} $\textbf{-synergy decomposition of the negentropy of the logical-AND gate.}}
		\label{tab:neg_and}
	\end{table}
	
	The first three states (all of which have $Y=0$) behave like the logical-XOR system. There is one bit of 1-synergy, and then all subsequent failures are indistinguishable from maximum entropy and so have 0 bit. The last state, $(1,1,1)$ has a more robust deviation from independence, with an additional bit of synergy lost when considering failing pairs of elements. Finally, there is -1 bit of partial synergy in the third order term. How do we interpret this? Once again, we will refer to the definition $\partial n_{3}(\textbf{x}) = \partial h^{\mathbb{U}}_{3}(\textbf{x}) - \partial h^{\mathbb{P}}_{3}(\textbf{x})$. The third-order partial synergistic entropies refer to that information that is lost when all three channels fail, but would survive the failure of any single channel or pair of channels. The negative partial synergistic negentropy value indicates that there is more information lost when we believe $\mathbb{P}(1,1,1)$ than when we believe $\mathbb{U}(1,1,1)$.
	
	\subsection*{Single-target mutual information}
	
	The original impetus for the partial information decomposition was the problem of decomposing the information that a set of inputs disclosed about a single target: $I(X_1,\ldots,X_k ; Y)$. Thus far, we have only considered un-directed statistics: the joint entropy, the Kullback-Leibler divergence, etc. However, it would be nice if the $\alpha$-synergy decomposition could also be brought to bear on the problem of directed information. This turns out to be straight-forward: once again following the logic of Varley \cite{varley_generalized_2024}, we can re-construct the directed mutual information from undirected Kullback-Leibler divergences. Recall that:
	
	\begin{align}
	\label{eq:mi_decomp_1}
	I(\textbf{X} ; Y) &= \mathbb{E}_{\mathbb{P}(Y)}\bigg[\sum_{\textbf{x}\in\boldsymbol{\mathcal{X}}}\mathbb{P}(\textbf{x}|y)\log\frac{\mathbb{P}(\textbf{x}|y)}{\mathbb{P}(\textbf{x})}\bigg]
	\end{align} 
	
	To decompose the information that \textbf{X} discloses about $Y$ then, requires doing the $\alpha$-synergy decomposition on the Kullback-Leibler divergence of $\mathbb{P}(\textbf{X}|y)$ from $\mathbb{P}(\textbf{X})$ for every $y\in\mathcal{Y}$. The resulting set of $\alpha$-synergy atoms gives the average amount of information that \textbf{X} discloses about $Y$ that would be lost if any single $X_i$ failed, if any pair $X_i, X_j$ failed, and so on for all $\alpha$ in $\{1,\ldots,k\}$. 
	
	Like the lattice-based decomposition of the Kullback-Leibler divergence, the $\alpha$-synergy decomposition of the mutual information inherits the unusual property that the number of atoms will change depending on exactly how the mutual information is defined \cite{varley_generalized_2024}. If we use the formulation given in Eq. \ref{eq:mi_decomp_1}, there will be as many atoms as there are elements of \textbf{X}. However, there is another, equivalent formulation of $I(\textbf{X};Y)$, also based on Kullback-Leibler divergences:
	
	\begin{equation}
	I(\textbf{X};Y) = \sum_{\textbf{x},y\in\boldsymbol{\mathcal{X}}\times\boldsymbol{\mathcal{Y}}}\mathbb{P}(\textbf{x},y)\log\frac{\mathbb{P}(\textbf{x},y)}{\mathbb{P}(\textbf{x})\mathbb{P}(y)}
	\end{equation}
	
	In this case, rather than returning an atom for every $X_i$ in \textbf{X}, it will return $|\textbf{X}|+1$ atoms, since in this formulation, the joint state of $\{\textbf{x},y\}$ is considered directly, rather than separated as in Eq. \ref{eq:mi_decomp_1}. Previously, Varley speculated as to the conceptual implications of this \cite{varley_generalized_2024} and interested readers are invited to visit the cited literature for further discussion.  
	
	One derivative of the partial information decomposition that does not have an obvious counterpart in the $\alpha$-synergy framework is the integrated information decomposition ($\Phi$ID) \cite{mediano_towards_2021}. The $\Phi$ID relaxes the single-target specification of the classic PID, allowing for the decomposition of the information shared by two sets of random variables: $I(\textbf{X};\textbf{Y})$. Currently, the $\Phi$ID is not describable as a Kullback-Leibler divergence, instead being based on a product-lattice structure, and cannot currently be reconstructed from the framework presented here. Future work on the $\alpha$-synergy decomposition should focus one finding a way to produce an analogue of the $\Phi$ID in the same way that the single-target decomposition above is analogous to the PID. 
	
	\section*{Extensions beyond information: Structural synergy}
	
	Thus far, we have focused on the information-theoretic measure of local entropy, and the various measures that can be constructed from sums and differences of it. However, the logic of the $\alpha$-synergy decomposition can be applied beyond the world of information theory to a larger family of functions on sets, so long as a few conditions are met. For a function $f()$ on a set \textbf{X} to induce an interpretable $\alpha$-synergy decomposition, I conjecture that it must satisfy a minimal set of properties:
	\begin{enumerate}
		\item \textit{Localizability:} If $f(\textbf{X})$ is an expected value, the $f()$ must also be defined on every local instance of $\textbf{X}=\textbf{x}$. 
		\item \textit{Symmetry:} $f(\textbf{x})$ is invariant under any permutation of the elements of \textbf{x}.
		\item \textit{Non-negativity:} $f(\textbf{x})) > 0$. 
		\item \textit{Monotonicity:} if $\textbf{y}\subseteq\textbf{x}$, then $f(\textbf{y}) \leq f(\textbf{x})$. 
	\end{enumerate} 
	
	If $f()$ satisfies these rules, then one can define an $\alpha$-synergy function on it:
	
	\begin{align}
	f_{\alpha}^{syn}(\textbf{x}) &= \min_{\substack{\textbf{a}\subset \{k\} \\ |\textbf{a}|=\alpha}}\bigg[f(\textbf{x}) - f(\textbf{x}^{-\textbf{a}})\bigg]
	\end{align} 
	
	and recover the same non-negative backbone decomposition as described above. This opens the door to the analysis of higher-order interactions beyond the information-theoretic perspective, and how the structure of the system changes as it disintegrates. 
	
	To demonstrate this, we will consider a classic family of models in complexity science: the bivariate network, and the question of communication over edges from node to node. There are many different strategies for communicating over a complex network \cite{seguin_brain_2023}. Generally, the ease with which a signal can be sent from an arbitrary source to an arbitrary target is considered as a measure of network ``integration" \cite{rubinov_complex_2010}, and given a sufficiently well-behaved measure (satisfying the desiderata detailed above), we can ask ``how much of the overall integration depends on the particular pattern of edges in the network?" We refer to this idea as the ``structural synergy." 
	
	The measure that we use here is the \textit{communicability} \cite{estrada_communicability_2009}, which measures how readily a signal radiating from a source node will reach all of the others, assuming unbiased, random diffusion. For an adjacency matrix \textbf{M}, the communicability is given by:
	
	\begin{equation}
	C(\textbf{M}) = \mathbb{E}\bigg[\big(e^{\textbf{M}}\big)_{ij}\bigg]
	\end{equation}
	
	Where the expectation is taken over all edges $\textbf{M}_{ij}$. The communicability can be readily localized: the integration between any two nodes $i$ and $j$ is just value of cell $(i,j)$ in the communicability matrix $e^{\textbf{M}}$. So, for every edge, it is possible to define the difference between the communicability between $i$ and $j$ when the network is unperturbed and the communicability between $i$ and $j$ after some set of edges has failed. The interpretation of the $\alpha$-synergy in this case is the same as in the information-theoretic case, although the specific measure is different. The 1-synergy is that integration that depends on the existence of all edges, and would be destroyed by the failure of any single edge. Likewise, the 2-synergy is the integrations that depends on the existence of all pairs of edges and so-on. 
 	
	\begin{figure}
		\centering 
		\includegraphics[width=\textwidth]{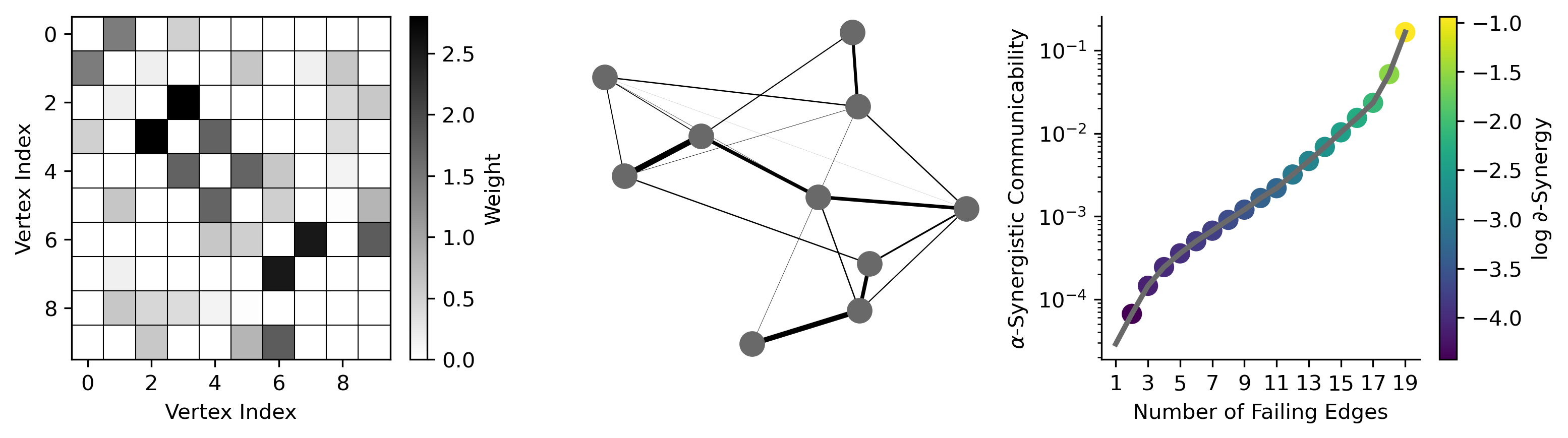}
		\caption{\textbf{Communicability example:} For a randomly generated Erdos-Renyi graph with ten nodes, and edge weights pulled from an exponential distribution with $\lambda=1$, we compute the $\alpha$-synergy spectrum for the communicability across all possible partitions of the edge set, and compute the $\alpha$-synergy and $\partial$-synergy for each one. On the leftmost panel is the weighted adjacency matrix, with the associated network in he middle panel. In the rightmost panel, we plot the expected $\alpha$-synergistic communicability against the number of failing edges, with the points coloured by the partial synergistic communicability.}
		\label{fig:commun}
	\end{figure}

	In Figure \ref{fig:commun} and example decomposition is demonstrated for a small Erdos-Renyi graph with ten nodes and nineteen edges. It is clear that there is very little synergistic communicability in this graph: the difference between the 1-synergy and the total communicability spans four orders of magnitude. This is unsurprising given that the communicability strategy inherently involves diffusing signals over every possible path - almost all edges have to fail before there's no way to get a signal from $i$ to $j$ due to the redundancy inherent in the measure. The structural synergy, while arguably negligible, is nevertheless calculable and present in the network - future work exploring how the topology of pairwise edges influences the higher-order synergies, and the significance for dynamics on the networks may reveal novel links between how the structure of lower-order causal mechanisms facilitates the emergence of higher-order statistical structures.  
	
	The structural synergy is one of a number of recent approaches for using information theory to explore emergent structures in graphs and networks that has been developed. Rosvall et al. used the notion of information compression as a framework for finding communities in complex networks (known as the Infomap algorithm) \cite{rosvall_maps_2008,rosvall_map_2009}. Later, Klein et al. formulated an approach for characterizing coarse-grained higher scales, taking an explicitly emergentist perspective \cite{klein_emergence_2020}, finding that some networks can be coarse-grained in such a way that the predictive information at the macro-scale is greater than the predictive information at the micro-scale, and that evolution seems to optimize for this property in biological interaction networks \cite{klein_evolution_2021}. Klein et al.,'s approach operationalizes the information-content of a network via a random-walker model (how does the uncertainty about the location of a walker on the network at time $t$ change upon learning the location of the walker at time $t-1$). Subsequently, Varley showed that the notion of coarse-graining emergence explored by Klein could be localized to individual edges \cite{varley_flickering_2023}, coining the term ``flickering emergence" to describe the local variability in emergent structure over the edges of the network, or through time in a dynamic process. Since the structural synergy can also be formulated with respect to random walker dynamics (e.g. the communicability), we propose that there may be links between the structural synergy and the effective emergence, although the precise nature of the relationship remains an area of future work. Recently, Luppi et al. proposed an approach by which the interactions of different layers of a multi-layer network could be understood in terms of redundant, unique, and synergistic interactions between shortest paths, which they term the partial network decomposition (PND) \cite{luppi_quantifying_2023}. The notion of structural synergy in the context of the $\alpha$-synergy decomposition is distinct from the notion of synergy in the PND for several reasons: the first is that the structural synergy is not specifically defined on shortest paths, but rather any functional graph invariant that satisfies the given desiderata (although we note that the efficiency, or the reciprocal of the shortest path length, would work). Second, the PND compares two networks defined on the same set of nodes, while the structural synergy characterizes a single network \textit{qua} itself. 
	
	In the context of networks, the notion of structural synergy is something of an inversion of the usual approach to analysing complex networks. Typically, analyses focus on individual nodes (first-order structures), or pairwise edge-centric perspectives \cite{ahn_link_2010,betzel_living_2023} (second order structures). In contrast, the structural synergy takes a top-down approach, describing the irreducible structure in the whole as a function of the joint-state of all of the parts.
	
	\subsection*{Discussion}
	
	In this work, we have introduced the general notion of a decomposition of synergistic ``structure" in complex systems that encompass both information-theoretic, and topological definitions of ``structure", and can be applied to a variety of further contexts. 
	
	Our approach was developed initially to address the limitations in commonly used, information-theoretic approaches for understanding statistical ``synergy". The two most commonly used approaches are the partial information decomposition \cite{williams_nonnegative_2010} (and subsequent derivatives \cite{ince_partial_2017,finn_generalised_2020,varley_generalized_2024}), and the O-information \cite{rosas_quantifying_2019} (and subsequent derivatives \cite{stramaglia_quantifying_2021,scagliarini_gradients_2023}). The partial information decomposition-based approach provides a ``complete" map of the structure of multivariate information in the form of the antichain lattice. However, the super-exponential growth of the lattice makes it impractical for all but the smallest toy systems. Furthermore, it is a ``redundancy-first" approach: synergistic information is implicitly defined in terms of redundancy, and different definitions of redundancy imply qualitatively and quantitatively different notions of synergy. In contrast, the O-information scales much more gracefully than the information decompositions, however it only reports whether redundancy or synergy dominates a system, and has a very conservative notion of ``synergy", analogous to the 1-synergy in our approach. 
	
	In summary, the PID is ``complete", but does not scale, and hinges on a measure of redundancy, while the O-information scales nicely and does not require defining redundancy, but does not provide a map of the structure of the system in question. The $\alpha$-synergy decomposition was designed to balance these trade-offs. It is a ``synergy-first" approach, hinging on a definition of synergy based on channel failures (somewhat like the dual total correlation \cite{abdallah_measure_2012}) and to scales more elegantly than the antichain lattice. Unlike the O-information, however, we get a measure of partial synergy for every scale, rather than just the top. the $\alpha$-synergy has its limitations too, however. The most significant is that it does not reference how synergy is distributed over the various elements and sets of elements, instead homogenizing the system under the summary statistics of the minimum/maximum/average functions: all of the structure is squished down onto the one-dimensional backbone. Furthermore, while it scales more elegantly than the PID, the complete $\alpha$-synergy decomposition still eventually becomes intractable, requiring optimizations that may break some of the mathematical guarantees that make it work. Whether a given scientist reaches for the PID/PED/GID, the O-information and related measures, or the $\alpha$-synergy decomposition will depend on the specific nature of the questions being asked, and care should be taken to ensure that the right approach is being used for a given analysis. 
	
	The measure that the $\alpha$-synergy decomposition has the most in common with is the Tononi-Sporns-Edelman complexity \cite{tononi_measure_1994}, which takes a similar approach to inferring the structure of a system by sweeping all possible bipartitions of the structure. The TSE complexity is based on the total correlation specifically, and explicitly compares the expected total correlation at each scale to a built-in null of maximum entropy. In contrast, the $\alpha$-synergy decomposition can be applied to any function that can be constructed from sums or differences of entropies (including the total correlation) and does not have the built-in null. Furthermore, in the TSE complexity, each subset is considered individually, while in our approach, it is the interaction between a subset and its complement that is significant (the term $h(\textbf{x}^\textbf{a}|\textbf{x}^{-\textbf{a}}))$. The $\alpha$-synergy also bears some similarity to the ``synergistic disclosure" decomposition from Rosas et al., \cite{rosas_operational_2020}. The synergistic disclosure framework is a synergy-first based approach (unlike the usual PID), however it still relies on the antichain lattice structure, and so suffers from the explosive growth in the number of terms. To address this, Rosas et al., provide a compressed, one-dimensional representation of the lattice (the ``backbone", from which our approach takes its name), with a very similar structure to the backbone presented here. The synergistic disclosure approach relies on a very different notion of synergy, however, one based on the logic of data privacy \cite{rassouli_data_2020} and is specific to the expected mutual information (i.e. it cannot be readily localized or extended to other measures like the Kullback-Leibler divergence or the total correlation). Finally, the $\alpha$-synergy decomposition shows some resemblances to the gradients of O-information recently introduced by Scalgiarini et al., \cite{scagliarini_gradients_2023}, which also considers how the information structure of a system changes in response to sets of failures, although like the TSE complexity, it depends on a particular measure (in this case, the O-information) rather than being a general approach. The gradients of O-information are also focused on finding how \textit{specific} elements, or low-order combinatons of elements contribute to higher-order information circuits, while the $\alpha$-synergy decomposition ignores node-specific information entirely to generate summary statistics. Finding a way to combine these approaches (gradients of $\alpha$-synergy) might help get the best of both worlds: a tractable picture of multi-scale synergy that also shows how it is represented over the specific elements. 
	
	One possible application of structural synergy is the study of cascading failures in complex systems and the inherent tradeoffs between robustness and efficiency. Following the COVID-19 pandemic, many global supply chains that relied on just-in-time logistics failed when disruptions to manufacturing and distribution compromised the precise timing necessary for highly-efficient organization to function \cite{olson_trade-offs_2014}. This could be seen as a case of structural synergy in action: just-in-time supply chains require every part of the system to be in the correct ``state" at the correct time. Any deviation from that state (analogous to a failure in our framework) compromises the required synergy and pushes the system towards a new configuration. Extending the structural synergy and $\alpha$-synergy decompositions to dynamical processes may help provide new insights into the design of efficient and robust systems. 
	
	\subsection*{Conclusions}
	
	In this paper, we have introduced the $\alpha$-synergy decomposition, a novel approach to measuring the synergistic information in a multivariate system based on the effects of random failures. The $\alpha$-synergy decomposition was designed to balance the tradeoffs associated with existing measures: it scales more gracefully than the partial information decomposition \cite{williams_nonnegative_2010} at the expense of losing element-specific information. Conversely, it is less scalable than the O-information \cite{rosas_quantifying_2019}, but it provides a ``spectrum" of synergies for each scale, rather than just the redundancy-synergy balance. From the decomposition of the entropy, it is possible to reconstruct almost all other formulations of the partial entropy decomposition, including the single-target mutual information and the generalized Kullback-Leibler decomposition. Excitingly, the same logic can be applied outside of the domain of information theory: we introduce the notion of ``structural synergy" and describe the contexts in which it can be applied. The structural synergy can be used to assess how the particular pattern of elements or interactions in a complex system contributes to some property of interest such as integration or productive capacity. The $\alpha$-synergy decomposition, being both practically useful and very general, may represent a significant, novel tool in the study of part-whole relationships in complex systems. 
	
	\section*{Acknowledgements}
	I would like Dr. Caio Seguin for extensive discussions about how multivariate information decomposition might be ported to the problem of network communication measures. I would also like to thank Ms. Maria Pope and Dr. Olaf Sporns for listening to me ramble about this idea over Zoom. Finally, I would like to thank Dr. Joshua Bongard for allowing me to pursue this work.
	

	\newpage
	
	\subsection*{Appendix}
	
	\subsubsection*{Proof 1: The minimum $\boldsymbol{\alpha}$-synergy function is monotonic}
	
	For the minimum $\alpha$-synergy function to be monotonic, we must show that, for two (potentially overlapping) subsets of \textbf{x}: \textbf{a} and \textbf{b}, with $|\textbf{a}|=\alpha$, $|\textbf{b}|=\beta$, and $\alpha<\beta$ that:
	
	\begin{align}
	\min_{\substack{\textbf{b}\subset \{k\} \\ |\textbf{b}|=\beta}}\bigg[h(\textbf{x}) - h(\textbf{x}^{-\textbf{b}})\bigg] \geq 
	\min_{\substack{\textbf{a}\subset \{k\} \\ |\textbf{a}|=\alpha}}\bigg[h(\textbf{x}) - h(\textbf{x}^{-\textbf{a}})\bigg]
	\end{align}
	
	Intuitively, this says that the minimally invasive perturbation on $\alpha$ nodes should destroy less information than the minimally invasive perturbation on $\beta$ nodes. 
	
	We will define the ``winning" subset of size $\alpha$ as $\textbf{a}^*$ and the ``winning" subset of size $\beta$ as $\textbf{b}^*$. Since $\beta > \alpha$, we can partition $\textbf{b}^*$ into two non-overlapping subsets \textbf{p} and \textbf{q}, with the stipulation that $|\textbf{p}|=\alpha$. By the definition of $\textbf{a}^*$ we know that:
	
	\begin{equation}
	\bigg[h(\textbf{x})-h(\textbf{x}^{-\textbf{p}})\bigg] \geq \bigg[h(\textbf{x})-h(\textbf{x}^{-\textbf{a}^*})\bigg]
	\end{equation}
	
	The loss of the elements in \textbf{p} destroys at least as much information as the loss of all elements in $\textbf{a}^*$. However, that still leaves all the elements in \textbf{q} unaccounted for. 
	
	Since $\textbf{x}^{-\{\textbf{p}\cup\textbf{q}\}}\subseteq\textbf{x}^{-\textbf{p}}$ it follows that $h(\textbf{x}^{-\{\textbf{p}\cup\textbf{q}\}}) \leq h(\textbf{x}^{-\textbf{p}})$, with equality if $\textbf{q}\subset\textbf{p}$, which cannot happen as $\textbf{q}\cap\textbf{p}=\emptyset$ by construction. This leaves us with the chain of inequalities:
	
	\begin{equation}
	\bigg[h(\textbf{x})-h(\textbf{x}^{-\{\textbf{p}\cup\textbf{q}\}})\bigg] > \bigg[h(\textbf{x})-h(\textbf{x}^{-\textbf{p}})\bigg] \geq \bigg[h(\textbf{x})-h(\textbf{x}^{-\textbf{a}^*})\bigg]
	\end{equation}
	
	By definition, $\{\textbf{p}\cup\textbf{q}\}=\textbf{b}^*$, so the statement: 
	
	\begin{equation}
	\bigg[h(\textbf{x})-h(\textbf{x}^{-\textbf{b}^*})\bigg] \geq \bigg[h(\textbf{x})-h(\textbf{x}^{-\textbf{a}^*})\bigg]
	\end{equation}
	
	is true. $\square$ 
	
	\subsubsection*{Proof 2: The maximum $\boldsymbol{\alpha}$-synergy function is monotonic}
	
	For the maximum $\alpha$-synergy function to be monotonic, we must show that, for two (potentially overlapping) subsets of \textbf{x}: \textbf{a} and \textbf{b}, with $|\textbf{a}|=\alpha$, $|\textbf{b}|=\beta$, and $\alpha<\beta$ that:
	
	\begin{align}
	\max_{\substack{\textbf{b}\subset \{k\} \\ |\textbf{b}|=\beta}}\bigg[h(\textbf{x}) - h(\textbf{x}^{-\textbf{b}})\bigg] \geq 
	\max_{\substack{\textbf{a}\subset \{k\} \\ |\textbf{a}|=\alpha}}\bigg[h(\textbf{x}) - h(\textbf{x}^{-\textbf{a}})\bigg]
	\end{align}
	
	Intuitively, this means that, when $\alpha$ elements fail, the worst-case scenario in terms of information loss must be less severe than the worst-case scenario when $\beta$ elements fail. The logic is similar to the logic in the above proof. 
	
	Let $\delta=\beta-\alpha$. Suppose $\textbf{a}^*$ is the winning subset that maximizes $h(\textbf{x})-h(\textbf{x}^{-\textbf{a}})$. We define a new set \textbf{d} of $\delta$ elements. The entropy of the union $h(\textbf{x}^{-\{\textbf{a}^*\cup \textbf{d}\}}) \leq h(\textbf{x}^{-\textbf{a}^*})$, and so there must be at least one set $\textbf{b}^*$ that has entropy less than $\textbf{a}^*$, giving the inequality:
	
	\begin{equation}
	h(\textbf{x}) - h(\textbf{x}^{-\textbf{b}^*}) \geq h(\textbf{x}) - h(\textbf{x}^{-\textbf{a}^*}) 
	\end{equation}
	
	The maximum $\alpha$-synergistic entropy function will find the largest of those $\textbf{b}^{*}$ by definition. $\square$
	
	\subsubsection*{Proof 3: The average $\boldsymbol{\alpha}$-synergy decomposition is monotonic}
	
	For the maximum $\alpha$-synergy function to be monotonic, we must show that, for two (potentially overlapping) subsets of \textbf{x}: \textbf{a} and \textbf{b}, with $|\textbf{a}|=\alpha$, $|\textbf{b}|=\beta$, and $\alpha<\beta$ that:
	
	\begin{align}
	h(\textbf{x}) - \mathbb{E}\big[h(\textbf{x}^{-\textbf{b}})
	\big] \geq h(\textbf{x}) - \mathbb{E}\big[h(\textbf{x}^{-\textbf{a}})\big] 
	\end{align}
	
	Where $\mathbb{E}\big[ h(\textbf{x}^{-\textbf{a}})\big]$ is the expected local entropy computed over all subsets \textbf{a} of size $\alpha$. The logic of this proof is very similar to the logic of Proof 1. For every possible set \textbf{b}, we can partition it into two, non-overlapping subsets \textbf{p} and \textbf{q}, with $|\textbf{p}|=\alpha$. Consequently $\mathbb{E}\big[h(\textbf{x}^{-\textbf{p}})\big] = \mathbb{E}\big[h(\textbf{x}^{-\textbf{a}})\big]$. Since $h(\textbf{x}^{-\{{\textbf{p}\cup \textbf{q}}\}})\leq h(\textbf{x}^{-\textbf{p}})$ for all \textbf{b}, then it follows that:
	
	\begin{equation}
	\mathbb{E}\big[h(\textbf{x}^{-\textbf{b}})
	\big] \leq \mathbb{E}\big[h(\textbf{x}^{-\textbf{a}})\big] 
	\end{equation}
	
	and that 
	
	\begin{align}
	h(\textbf{x}) - \mathbb{E}\big[h(\textbf{x}^{-\textbf{b}})
	\big] \geq h(\textbf{x}) - \mathbb{E}\big[h(\textbf{x}^{-\textbf{a}})\big] 
	\end{align}
	$\square$

\end{document}